\documentclass[format=sigconf, review=false, screen=true, table, dvipsnames, preprint=true]{acmart}

\usepackage[utf8]{inputenx}
\input{ix-utf8enc.dfu}
\usepackage[T1]{fontenc}

\usepackage{csquotes}
\usepackage[english]{babel}

\usepackage{subcaption}
\usepackage[flushleft]{threeparttable}
\usepackage{tabu, tabularx, booktabs}
\newcolumntype{Y}{>{\arraybackslash}X}
\newcolumntype{Z}{>{\columncolor{Gray!25}\arraybackslash}X}

\makeatletter
\newcommand\threepart@subtable{
  \caption@setoptions{threepartsubtable}%
  \caption@ORI@threeparttable
}
\newenvironment{threepartsubtable}{%
  \threepart@subtable
}{%
  \endthreeparttable
}
\makeatother
\usepackage{graphicx}

\graphicspath{{figures/}}

\makeatletter
\define@key{Gin}{resolution}{\pdfimageresolution=#1\relax}
\makeatother

\usepackage{cleveref}
\crefname{figure}{figure}{figures}
\crefname{equation}{eq.}{eqs.}

\usepackage{amsmath}
\usepackage{amssymb}
\usepackage{mathrsfs}
\usepackage{mathtools}
\mathtoolsset{centercolon}
\usepackage{mathbbol}
\DeclareMathAlphabet{\mathcal}{OMS}{cmsy}{m}{n}

\makeatletter
\DeclareRobustCommand*\cal{\@fontswitch\relax\mathcal}
\makeatother

\usepackage[all]{foreign}
\usepackage{xspace, xpunctuate}

\mathchardef\mhyphen="2D

\newcommand{\rfrac}[2]{{}^{#1}\!/_{#2}}

\DeclareRobustCommand\textastdbl{%
	\leavevmode
	{\sbox0{\ddag}%
		\ooalign{\raisebox{\ht0-\height}{\small*}\cr
			\raisebox{\depth-\dp0}{\scalebox{1}[-1]{\small*}}\cr}%
	}%
}
\def\wl{\textsuperscript{\small\textasteriskcentered}}
\def\bl{\textsuperscript{\textastdbl}}
\def\wc{\textsuperscript{\dag}}
\def\bc{\textsuperscript{\ddag}}

\newcommand{\ERM}{\ensuremath{RM_E}\xspace}
\newcommand{\TERM}{\ensuremath{RMT_{E}}\xspace}
\newcommand{\EKDE}{\ensuremath{KDE_E}\xspace}
\newcommand{\TERMEKDE}{\ensuremath{KDE{+}KDE_E{+}RMT_{E}}\xspace}

\newcommand{\selected}[1]{\CollectionSet_{#1}}
\newcommand{\numselected}[1]{{\vert \selected{#1} \vert}}

\newcommand{\VerticalSet}{{\cal C}}
\newcommand{\VerticalSetSize}{{\vert \VerticalSet \vert}}
\newcommand{\vertical}{c}
\newcommand{\verticalseqlength}{{\VerticalSetSize}}
\newcommand{\verticalseq}{\vertical_1 \vertical_2 \cdots \vertical_\verticalseqlength}

\newcommand{\ExternalCollection}{\CollectionSet}
\newcommand{\CollectionSet}{{\cal C}}

\newcommand{\Collection}{c}

\newcommand{\InstantSet}{T}
\newcommand{\instant}{{\MakeTextLowercase{\InstantSet}}}

\newcommand{\InstantWeightSet}{\lambda}
\newcommand{\instantweight}{{\MakeTextLowercase{\InstantWeightSet}}}

\newcommand{\maximumlikelihoodLow}{maximum-likelihood\xspace}

\newcommand{\querylikelihoodLow}{query-likelihood\xspace}

\newcommand{\prfLow}{pseudo-relevance feedback\xspace}

\newcommand{\learningtorankCap}{Learning to rank\xspace}
\newcommand{\learningtorankCAP}{Learning to Rank\xspace}
\newcommand{\learningtorankLow}{learning to rank\xspace}

\newcommand{\kmeansCap}{k-Means\xspace}

\newcommand{\kernelsym}{K}
\newcommand{\kernel}{\kernelsym \left( z \right)}
\newcommand{\kernelBandwidth}{h}
\newcommand{\optimalKernelBandwidth}{\kernelBandwidth^*}
\newcommand{\gaussiansym}{\mathcal{N}}
\newcommand{\gaussianKernel}{\gaussiansym \left( z, 0 \right)}

\newcommand{\QuerySet}{Q}

\newcommand{\query}{{\MakeTextLowercase{\QuerySet}}}

\newcommand{\origQuery}{\query}
\newcommand{\origQueryModel}{\langModel_{\origQuery}}

\newcommand{\extExpQueryModel}{\langModel_{F_\CollectionSet}}
\newcommand{\finalQuery}{\query'}
\newcommand{\finalQueryModel}{\langModel_{\finalQuery}}

\newcommand{\timeperiod}{{T}}

\newcommand{\RetrievedSet}{{\cal R}}

\newcommand{\RetrievedSetCollection}{{\RetrievedSet_\Collection}}
\newcommand{\RetrievedSetCollectionSize}{\vert \RetrievedSetCollection \vert}

\newcommand{\doc}{d}

\newcommand{\docseqlength}{k}
\newcommand{\docseq}{\doc_1 \cdots \doc_\docseqlength}

\newcommand{\doccontent}{\word_\doc}
\newcommand{\docmeta}{m_\doc}
\newcommand{\doctime}{\instant_\doc}

\newcommand{\corpus}{D}
\newcommand{\word}{w}

\newcommand{\cnt}[2]{{\rm \#}(#1, #2)}

\newcommand{\queryLength}{\vert \query \vert}

\newcommand{\langModel}{{\MakeTextLowercase{\theta}}}

\newcommand{\set}[1]{\{#1\}}

\newcommand{\corpussize}{N}

\newcommand{\relvar}{R}

\newcommand{\probsym}{P}
\newcommand{\prob}[1]{\probsym(#1)}
\newcommand{\condprobgen}[3]{#1(#2 \mid #3)}
\newcommand{\condprob}[2]{\condprobgen{\probsym}{#1}{#2}}

\newcommand{\undercondprobgen}[4]{#1_#2(#3 \mid #4)}
\newcommand{\undercondprob}[3]{\undercondprobgen{\probsym}{#1}{#2}{#3}}

\begin{document}

\copyrightyear{2019} 
\acmYear{2019} 
\setcopyright{acmlicensed}
\acmConference[WSDM '19]{The Twelfth ACM International Conference on Web Search and Data Mining}{February 11--15, 2019}{Melbourne, VIC, Australia}
\acmBooktitle{The Twelfth ACM International Conference on Web Search and Data Mining (WSDM '19), February 11--15, 2019, Melbourne, VIC, Australia}
\acmPrice{15.00}
\acmDOI{10.1145/3289600.3290966}
\acmISBN{978-1-4503-5940-5/19/02}

\settopmatter{printacmref=true}

\title{Modeling Temporal Evidence from External Collections\textsuperscript{*}}
\thanks{* Please cite the WSDM 2019 version of this paper.}

\author{Flávio Martins}
\affiliation{
    \institution{NOVA LINCS}
    \institution{School of Science and Technology}
    \institution{Universidade NOVA de Lisboa}
    \postcode{2829-516}
    \city{Caparica}
    \country{Portugal}
}
\email{flaviomartins@acm.org}
\author{João Magalhães}
\affiliation{
    \institution{NOVA LINCS}
    \institution{School of Science and Technology}
    \institution{Universidade NOVA de Lisboa}
    \postcode{2829-516}
    \city{Caparica}
    \country{Portugal}
}
\email{jm.magalhaes@fct.unl.pt}
\author{Jamie Callan}
\affiliation{
    \institution{Language Technologies Institute}
    \institution{School of Computer Science}
    \institution{Carnegie Mellon University}
    \streetaddress{5000 Forbes Avenue}
    \city{Pittsburgh} 
    \state{PA} 
    \postcode{15213}
    \country{USA}
}
\email{callan@cs.cmu.edu}

\renewcommand{\shortauthors}{F. Martins et al.}

\begin{abstract}
Newsworthy events are broadcast through multiple mediums and prompt the crowds to produce comments on social media.
In this paper, we propose to leverage on this behavioral dynamics to estimate the most relevant time periods for an event (\ie query).
Recent advances have shown how to improve the estimation of the temporal relevance of such topics. 
In this approach, we build on two major novelties. 
First, we mine temporal evidences from \textit{hundreds of external sources} into topic-based external collections to improve the robustness of the detection of relevant time periods.
Second, we propose a formal retrieval model that \textit{generalizes the use of the temporal dimension} across different aspects of the retrieval process.
In particular, we show that temporal evidence of external collections can be used to (i) infer a topic's temporal relevance, (ii) select the query expansion terms, and (iii) re-rank the final results for improved precision.
Experiments with TREC Microblog collections show that the proposed time-aware retrieval model makes an effective and
extensive
use of the temporal dimension to improve search results over the most recent temporal models.
Interestingly, we observe a strong correlation between precision and the temporal distribution of retrieved and relevant documents.
\end{abstract}

\maketitle

\section{Introduction}
A networked world and the increasing pervasiveness of Internet access enables the rapid adoption of new online communication mediums to discuss current events.
Previous research has explored this symbiosis between Twitter and the news~\cite{kwak_what_2010,sankaranarayanan_twitterstand_2009} and linked the two mediums~\cite{tsagkias_linking_2011,guo_linking_2013}.
Events are discussed on the Web as they happen and people following them can add to the conversation immediately.
Hence, improving the \textit{temporal relevance estimation} for searching such events became a significant research priority.

Nowadays, the state-of-the-art Web search systems are based on \learningtorankLow feature models that combine multiple text retrieval functions as well as other features.
Relevance on Twitter has many dimensions:
authority, popularity, freshness, geographical context, and topical relevance.
Previously, time-aware ranking research explored the assumption that fresh documents are more relevant~\cite{li_timebased_2003}. 
Later models revised this assumption in line with what is observed in Twitter: for time-sensitive queries, documents tend to cluster temporally~\cite{dakka_answering_2012, efron_temporal_2014}. 
Our approach is based on the intuition that discussions about a topic and its subtopics are likely to occur around the same time across multiple mediums.

The rationale is that newsworthy events trigger a cascade of activity on the Web and Twitter.
This information can be useful for ranking and, in some cases, can be gathered with ease.
The news often have a good coverage of current topics, clean journalistic language, and reliable timestamps.
Thus, it is desirable to mine news sources to offer more context to the \emph{tweets} as well as to the users' queries intent. In particular, we aim to explore the \emph{crowd aggregation effect} to extract temporal evidence from news verticals.
Temporal evidence is further used to refine the selection of query expansion terms and to estimate query topics temporal relevance.
This approach is completed with the re-ranking of the final search results leading to improved precision.
Hence, the proposed method brings a series of novel contributions: 

\begin{itemize}
    \item Explore the crowd effect by aggregating posts published by news sources into topic-based external collections;

    \item Mining of crowds' temporal evidence at different granularities (\ie \textit{verticals}, \textit{documents}, and \textit{terms});

    \item A formal time-aware ranking model that unifies multiple temporal features into a single comprehensive retrieval model.

\end{itemize}

Including the temporal dimension at the different steps of the search engine pipeline, improves the accuracy of several retrieval tasks, leading to greater overall gains. This is possible because, the temporal dimension introduces stronger evidence in many decision tasks (\eg selection of query expansion terms).
Evaluation on the TREC 2013 and TREC 2014 Microblog Track datasets shows that the proposed retrieval model outperforms state-of-the-art methods.

This paper is organized as follows: in \Cref{sec:relatedwork} we present the related work; in \Cref{sec:formalmodel} the formal temporal ranking model is detailed and the following sections detail its implementation; evaluation is presented in \Cref{sec:study};  and a more fine-grained discussion of results in \Cref{sec:prvtm-results}.

\section{Related Work}
\label{sec:relatedwork}

In the past, several authors have proposed to use multiple collections to improve search results.
\citet{bendersky_effective_2012} point to the limitation that standard query formulation tasks such as term weighting and query expansion often use a single source of information.
In their query expansion experiments, they combined multiple information sources
from newswire and web corpora and found better retrieval effectiveness
than when using a single source of information.
\citet{weerkamp_exploiting_2012} developed a novel query modeling framework to combine evidence from multiple external collections.
An interesting property of this model is that, if we assume that the query-dependent collection importance $\condprob{\query}{\Collection}$ is uniformly distributed and that the importance of a document in the collection is $\condprob{\doc}{\Collection} = 1 / \RetrievedSetCollectionSize$,
we arrive at the formulation of Mixture of Relevance Models (MoRM) proposed by \citet{diaz_improving_2006}.

Several time-based \prfLow methods were proposed for retrieval in time-sensitive collections using the relevance modeling framework.
\citet{keikha_time-based_2011} proposed time-based relevance models 
where they assume that the publishing date has an effect on the terms.
They introduce a generative model of the query that
first selects a date and then a term based on the time and query.
They found this approach was able to improve the coverage of the expanded query over the different subtopics by using temporal information to weight and select expansion terms.
\citet{choi_temporal_2012} extended the framework proposed by \citet{keikha_time-based_2011}
by making a simplifying assumption that $\condprob{\doc}{\timeperiod,\query}$ can be equal to
$\condprob{\doc}{\query}$ since the temporal dimension is incorporated already in choosing $\doc$.
In this formulation, a relevance model for each time period is estimated using the retrieved documents published in time the time period $\timeperiod$.
Each of the relevance models are then weighted by $\condprob{\timeperiod}{\query}$ to obtain the final expansion terms over all the time periods.
\citet{arguello_learning_2011} also explored the use of external collections, in the form of verticals, to leverage first-order statistics from verticals to improve search results.
The method proposed in this paper also explores multiple collections (\ie organized into verticals) but moves far beyond first-order statistics.

Recent time-dependent ranking approaches have resorted to \learningtorankLow techniques~\citep{metzler_structured_2012, kanhabua_learning_2012, costa_learning_2014} and temporal query detection~\citep{dai_learning_2011} that exploit non-temporal and temporal features.
\citet{dai_learning_2011} propose to run each query against a set
of rankers, which are weighted based on the temporal profile of a query, and therefore minimize the risk of degraded performance due to misclassifying the query in terms of recency intent.
\citet{metzler_structured_2012} defined \textit{microblog event retrieval}
as a search task that goes beyond ad hoc retrieval.
To uncover subtopics from these streams of very short and noisy posts
they proposed a temporal query expansion technique.
Their technique divides the timeline into time spans of one hour,
and ranks them according to the proportion of messages posted during
the time spans that match the query.
The \textit{burstiness score} weights terms, so that when counts for the occurrences
of terms are higher than usual, these terms will have higher weights.

Microblog retrieval has very specific and time-pressed requirements.
For instance, \citet{jones_temporal_2007} note that queries that favor recency are just a subset of the time-sensitive queries.
Along this vein of thought, several authors \cite{peetz_using_2013, whiting_temporal_2012} proposed to leverage the temporal distribution of the pseudo-relevant documents.
\citet{whiting_temporal_2012} combines pseudo-relevant document term distribution and temporal collection evidence using a variant of PageRank over a weighted graph that models the temporal correlation between n-grams.
Another approach by \citet{peetz_using_2013} leverages the temporal distribution of the pseudo-relevant documents themselves.
For each query, bursty time periods are identified and then documents from these periods are selected for feedback. 
The query model is updated with new terms sampled from the higher quality documents selected.
\citet{dakka_answering_2012} evaluated a time-sensitive pseudo-relevance feedback method on manually selected topic subsets from various TREC collections (\eg TREC News Archive, TREC Time-sensitive Queries).
\citet{dakka_answering_2012} identified the need to find the important time periods for time-sensitive queries and to integrate temporal relevance in the ranking model. 
Their ranking model explicitly splits the lexical and temporal evidence in the documents: $\doccontent$, the words in the document and $\doctime$, the document's timestamp:
$\condprob{\doc}{\query} \propto \condprob{\doccontent}{\query} \cdot \condprob{\doctime}{\query}$.
They propose techniques to estimate the $\condprob{\doctime}{\query}$ using histograms, however this method might be cumbersome  as the calculation of histogram bins is linked to many parameters.

Recently, modeling temporal relevance was shown to be effective for searching time-sensitive collections.
\citet{craveiro_query_2014} explored the segmentation of textual news articles, so that it can be leveraged in the query expansion process to focus the expansion terms temporally.
\citet{efron_temporal_2014} proposed a general and principled retrieval model for microblog search with temporal feedback. Their approach models the temporal density of a query $\condprob{\instant}{\query}$ with a kernel density estimate, with all the advantages brought by this method: the natural smoothness of the resulting function and a fully automated way to estimate the model variables (\eg bandwidth selection is data-driven, a function of the initial rank). This estimated temporal relevance is then employed to re-rank documents with a log-linear model.
\citet{martins_barbara_2016} achieved state-of-the-art results when using multiple external sources such as Wikipedia edits, views and newswire articles. 
Following the same rationale, that term expansions should be biased
to draw from documents from relevant (bursty) time periods.
\citet{rao_temporal_2016} proposed capturing these by estimating
the parameters of a continuous hidden Markov model
that best explains the sequential dependencies in the temporal distribution
of documents retrieved in the initial feedback step,
computing the most likely state sequence using the Viterbi algorithm,
and drawing terms only from bursty states.
These findings have found their way into the architecture of search indexes.
\citet{wang_partitioning_2017} examined how the main index collection could be partitioned (\ie by day, by source, \etc). This provides the important insight that collections can actually be partitioned over time.

In contrast to previous work, we propose to use multiple news verticals to robustly identify the relevant time periods for each query, instead of relying only on the temporal distribution of pseudo-relevant documents \citep{dakka_answering_2012} or first-order statistics from verticals \citep{arguello_learning_2011}.

\section{Modeling Temporal Evidence from External Collections}
\label{sec:formalmodel}
Consider a retrieval corpus containing $\corpussize$ documents, represented by $\corpus$. 
To integrate the temporal relevance component in the ranking model \citet{dakka_answering_2012} decomposed the document in two different parts:
lexical evidence, the words in the document ($\doccontent$),
and temporal evidence, the document's timestamp ($\doctime$).
We consider an augmented ranking model that contemplates query-independent signals or metadata from the document, $\docmeta$, in addition to the lexical and temporal evidence as follows:
\begin{align}
\begin{split}
\condprob{\doc}{\query}
&= \condprob{\doccontent, \doctime, \docmeta}{\query}
\\
& \propto \condprob{\doccontent}{\query} \cdot \condprob{\doctime}{\query} \cdot \condprob{\docmeta}{\query}
\\
& \propto \underbrace{\condprob{\query}{\doccontent} \cdot \prob{\doccontent}}_{\text{\querylikelihoodLow model}} \cdot \condprob{\doctime}{\query} \cdot \prob{\docmeta}
\end{split}
\label{eq:initial_ranking_model}
\end{align}
where the final formulation follows from the two following  steps:
First, by applying the Bayes' rule to $\condprob{\doc}{\query}$ and eliminating the quotient $\prob{\query}$ based on the rank equivalence to get the well-known \querylikelihoodLow retrieval model.
Second, by assuming the independence between document metadata and the query, $\prob{\docmeta}$ can be taken as the query-independent importance of the document.

To instantiate the ranking model from \Cref{eq:initial_ranking_model}, we need to estimate three components: lexical, temporal, and query-independent.
The lexical component can be estimated using \emph{relevance models} (see \Cref{sec:etbrm}) or standard \querylikelihoodLow, where we assume that $\prob{\doccontent}$ is uniform.
In this paper, we focus on estimating the temporal component using external collections (see \Cref{sec:etr}).
The query-independent component can be estimated using values extracted from the metadata of the document (see \Cref{tab:prvtm-non-temporal-features}).

To estimate the temporal component, former models~\citep{efron_temporal_2014, dakka_answering_2012} assume that relevant temporal information is only available from the search corpus itself, $\corpus$, for instance via the temporal distribution of an initial set of feedback documents.
However,
temporal feedback on the corpus alone can be boosted by external sources~\citep{martins_barbara_2016}.
Therefore, we propose improving the estimation of temporal relevance using external collections, in addition to the retrieval corpus:
\begin{align}
\begin{split}
\condprob{\doctime}{\query}
&= \condprob{\doctime}{\query, \corpus, \VerticalSet} \\
&=\condprob{\doctime}{\query, \corpus} \cdot \condprob{\doctime}{\query, \VerticalSet},
\label{eq:temp_corpus_times_external}
\end{split}
\end{align}
where the last step follows if we assume that temporal evidence can be extracted from the search corpus $\corpus$ and from the external collections $\VerticalSet = \set{\verticalseq}$ independently.
The first part can be estimated from the temporal distribution of feedback documents retrieved using the query $\query$.
We calculate the temporal relevance according to the external collections as described in \Cref{sec:etr}.

We also propose to generate query expansions to improve the document ranking (\ie the lexical component) by leveraging the external collections to estimate time-based \emph{relevance models}.
In \Cref{sec:etbrm}, we present a novel external time-based relevance model to generate expanded query models for retrieval in the corpus.
The expanded query model is computed by taking into account lexical as well as temporal evidence contained in the external collections.

\subsection{External Temporal Relevance}
\label{sec:etr}

For a given query, different collections yield different temporal relevance estimates (\ie different probability distributions of relevance over time).
Therefore, we need to extend~\Cref{eq:temp_corpus_times_external} to combine all the different temporal relevance estimates from each external collection into a single robust estimate.
In our approach, we combine them using a weighted mixture of probability distributions
\begin{align}
\begin{split}
\condprob{\doctime}{\query, \VerticalSet}
&\propto \sum_{\Collection \in \CollectionSet} \condprob{\doctime}{\query,\Collection} \cdot \condprob{\Collection}{\query} \\
&\propto \sum_{\Collection \in \CollectionSet} \condprob{\doctime}{\query,\Collection} \cdot \condprob{\query}{\Collection} \cdot \prob{\Collection}
,
\end{split}
\end{align}
where $\condprob{\doctime}{\query,\Collection}$ is the importance of time $\doctime$ for the query $\query$ in the collection $\Collection$, $\condprob{\query}{\Collection}$ is the relevance of the collection $\Collection$ to the query $\query$, and $\prob{\Collection}$ is the query-independent collection prior.

Considering that we may have many external collections,
the calculation of temporal relevance over all of them raises efficiency concerns.
To solve this problem we follow \emph{federated search} research~\cite{shokouhi_federated_2011, martins_vertical_2018}, and
consider that only a few collections contain most of the temporal evidence for a given query $\query$.
Therefore, we can use only those collections to provide an adequate approximation
\begin{align}
\condprob{\doctime}{\query, \VerticalSet}
\propto \sum_{\Collection \in \selected{\query}} \condprob{\doctime}{\query,\Collection} \cdot \condprob{\query}{\Collection} \cdot \prob{\Collection}
,
\end{align}
where $\selected{\query}$ is a ranking of the most relevant collections to query $\query$, and the query-independent prior of the collection is considered uniform $\prob{\Collection} = 1 / \numselected{\query}$

To estimate the relevance of each collection $\Collection$ for a query $\query$, represented by $\condprob{\query}{\Collection}$, we consider a similar approach to the ReDDE resource selection algorithm~\citep{si_relevant_2003}.
Considering $M_k$, the final single ranking obtained by merging all the results retrieved from the selected collections $\selected{\query}$,
the relevance of collection $\Collection$ is given by the ratio between the number of its documents that make it into the top ranking, $M_c$, by the total documents retrieved, $M_k$:
\begin{equation}
    \condprob{\query}{\Collection} = \frac{\vert M_\Collection \vert}{\vert M_k \vert}
\label{eq:crel}
\end{equation}
There are several other options that can be used to estimate $\condprob{\query}{\Collection}$, they include other resource selection algorithms~\citep{si_relevant_2003,shokouhi_central-rank-based_2007,kulkarni_shard_2012,weerkamp_exploiting_2012,aly_taily_2013}.

\subsubsection{Vertical Temporal Feedback}
\label{sec:temporal_vertical_feedback}

For each document $\doc$ we would like to find 
$\condprob{\doctime}{\query,\vertical}$,
the probability of relevance of its timestamp $\doctime$
according to vertical $\vertical$ 
and the query $\query$. This probability follows the joint distribution $f_\vertical(\doctime)$,
\begin{equation}
\condprob{\doctime}{\query, \vertical} \sim f_\vertical(\doctime).
\end{equation}
Following \citet{efron_temporal_2014}, we estimate the probability density function $f_\vertical(\doctime)$ 
by learning the distribution of feedback documents using a weighted kernel density estimation method:
\begin{align}
f_\vertical(\instant) = \frac{1}{nh} \sum_{d \in \RetrievedSet_\vertical} \instantweight_\doc \,\, K \left( \frac{\instant - \doctime}{h} \right)
\end{align}
where $\instant$ is the timestamp of the input document, $\RetrievedSet_\vertical$ is the set of retrieved documents from the collection $\Collection$ and $\doctime$ corresponds to these documents' timestamps.
The kernel function $\kernel$ corresponds to the Gaussian kernel 
$\gaussianKernel$, and the optimal bandwidth can be 
estimated by a data-driven method such as Silverman's rule-of-thumb $\optimalKernelBandwidth 
\approxeq 1.06\, \sigma\, n^{-\rfrac{1}{5}}$.
Finally, 
$\instantweight_\doc$,
is a non-negative weight on timestamp
$\doctime$,
to weight each timestamp by its importance.
The weight $\instantweight_\doc$ of each document's timestamp is based on its relevance to the query, for instance, the document's \querylikelihoodLow model retrieval score or estimated from its position in the rank.

\subsection{External Time-based Relevance Models}
\label{sec:etbrm}

Relevance models provide a framework for term selection and estimation of the importance of terms for query expansion~\citep{lavrenko_relevance_2001}.
We propose to estimate relevance models and generate a final query $\finalQuery$ using external collections $\CollectionSet$, leveraging their temporal evidence,
\begin{align}
\condprob{\query}{\doccontent, \CollectionSet} \approx \condprob{\finalQuery}{\doccontent}.
\end{align}
Let $\origQueryModel$ be the original query model and $\extExpQueryModel$ an estimated feedback query model based on feedback documents $\docseq$ from multiple external collections.
Inspired by \citet{zhai_modelbased_2001}, the final query model is $\finalQueryModel = (1 - \alpha) \,\, \origQueryModel + \alpha \,\, \extExpQueryModel$.
In this formulation, the final query is a linear combination of the original query model, $\condprob{\word}{\origQueryModel}$, and the estimated feedback query model, $\condprob{\word}{\extExpQueryModel}$, using external collections:
\begin{align}
\condprob{\word}{\finalQueryModel}
    &= \lambda \cdot \condprob{\word}{\origQueryModel} + (1 - \lambda) \cdot \condprob{\word}{\extExpQueryModel},
\end{align}
where the original query is modeled using its \maximumlikelihoodLow estimate $\condprob{\word}{\origQueryModel} = \cnt{\word}{\query} / \queryLength$. 
Time is introduced in the second parcel of the above expression to improve the estimation of the feedback query expansion terms.
To this end, we integrate temporal feedback into term selection to make it time-aware.

We start by estimating the feedback query model by leveraging pseudo-relevant documents from multiple external collections using a formulation proposed by~\citet{weerkamp_exploiting_2012}:
\begin{align}
\begin{split}
\condprob{\word}{\extExpQueryModel}
&\propto \sum_{\Collection \in \CollectionSet} \condprob{\word}{\query, \Collection} \cdot \condprob{\Collection}{\query} \\
&\propto \sum_{\Collection \in \CollectionSet} \condprob{\Collection}{\query} \sum_{\doc \in \RetrievedSet_\Collection} \condprob{\word}{\doc,\query} \cdot \condprob{\doc}{\Collection}
\end{split}
\end{align}
where we limit the computation of $\condprob{\word}{\query, \Collection}$ to the top documents retrieved from each individual collection, $\RetrievedSetCollection$.
Furthermore, if we consider $\condprob{\doc}{\Collection}$ to be uniform (\ie equal to $1/\RetrievedSet_\vertical$), we obtain the following formulation:
\begin{align}
&\propto \sum_{\Collection \in \CollectionSet} \condprob{\Collection}{\query} \,\, \frac{1}{\RetrievedSetCollectionSize} \sum_{\doc \in \RetrievedSet_\Collection} \condprob{\word}{\doc} \cdot \condprob{\query}{\doc}
\end{align}
In this formulation, term selection is blind to the temporal dimension because time has no influence on the importance of expansion terms.
However, in time-sensitive collections, the words in documents published on relevant time periods are more important and therefore should have a higher weight in the final expanded query.
Therefore, the formulation above is modified by considering both lexical and temporal components of the documents:
\begin{align}
\propto \sum_{\Collection \in \CollectionSet} &\condprob{\Collection}{\query} \,\, \frac{1}{\RetrievedSetCollectionSize} \sum_{\doc \in \RetrievedSet_\Collection} \condprob{\word}{\doccontent,\doctime} \cdot \condprob{\query}{\doccontent,\doctime} \\
\approx \sum_{\Collection \in \selected{\query}} &\condprob{\Collection}{\query} \,\, \frac{1}{\RetrievedSetCollectionSize} \sum_{\doc \in \RetrievedSet_\Collection} \condprob{\word}{\doccontent} \cdot \condprob{\query}{\doccontent} \cdot \condprob{\doctime}{\query}
\end{align}
the last step stems from the fact that the probability of word $\word$ for document $\doc$ depends only on the document content $\doccontent$ and is independent of its timestamp $\doctime$, and that the probability of the query $\prob{\query}$ is constant.
Hence, this formulation assumes that $\condprob{\doctime}{\query}$ can be estimated from each document timestamp $\doctime$. As in~\Cref{sec:temporal_vertical_feedback}, we use kernel density estimation \cite{efron_temporal_2014} to provide a smooth estimate of $\condprob{\doctime}{\query}$.

Finally, an approximate relevance model is calculated using only the most relevant collections $\selected{\query}$ to the query $\query$, since they should contribute the most to the estimation of the final expansion term weights.
As in \Cref{eq:crel}, we assume $\prob{\Collection}$ to be constant and uniform.

\subsection{Estimation over Discrete Time Periods}

In the previous section, we proposed a method that assumes a continuous approach to the temporal dimension of relevance, using kernel density estimation to predict the importance of specific points in time.
However, it is linked to this specific estimation method while previous methods of estimating temporal relevance exist (\eg volume-based, histogram-based, window-based) and new methods will be proposed in the future.
Therefore, in this section, we discuss a general generative model of the query that relies instead on a discrete partitioning of a timeline into time periods.
For each $\word$, it first selects a collection, then a time period, and then a term based on the collection, time period, and the query. Formally,
\begin{align}
\begin{split}
\condprob{\word}{\extExpQueryModel}
&= \sum_{\Collection \in \CollectionSet} \sum_{\timeperiod} \condprob{\word}{\timeperiod,\query,\Collection} \cdot \condprob{\Collection}{\timeperiod,\query} \cdot \condprob{\timeperiod}{\query}\\
&\propto \sum_{\Collection \in \CollectionSet} \condprob{\Collection}{\query} \sum_{\timeperiod} \condprob{\word}{\timeperiod, \query,\Collection} \cdot \condprob{\timeperiod}{\query},
\end{split}
\end{align}
where $\condprob{\word}{\timeperiod,\query, \Collection}$ is the importance of the word $\word$ in time period $\timeperiod$ (\eg day, hour) for the query $\query$ given collection $\Collection$, $\condprob{\Collection}{\timeperiod,\query}$ is the importance of collection $\Collection$ in the time period $\timeperiod$ for the query $\query$, and $\condprob{\timeperiod}{\query}$ is the importance of the time period $\timeperiod$ to the $\query$.
The last step follows if we assume the importance of a collection to a query to be independent from any given time period, $\condprob{\Collection}{\timeperiod,\query} = \condprob{\Collection}{\query}$.

A similar deduction to what was followed in the previous section, leads to a discrete model, which is more compatible with previous research in temporal information retrieval.

\section{\learningtorankCAP Model Using External Temporal Evidence}
\label{sec:prvtm-rankingmodel}
We are now ready to plug-in the temporal evidence and the time-based relevance models from multiple verticals into a common ranking model. To combine the different temporal features extracted from multiple query-specific verticals, we first re-write \Cref{eq:initial_ranking_model} as the following log-linear model 
\begin{align}
\log \condprob{\doc}{\query} \propto Z + \log \condprob{\query}{\doccontent} + \log \condprob{\doctime}{\query} + \log \prob{\docmeta},
\label{eq:initial_log_ranking_model}
\end{align}
where we can replace $\condprob{\doctime}{\query}$ by the temporal relevance over $\corpus$ and $\ExternalCollection$, and the query $\query$ by the expanded query $\finalQuery$.
Then we can use \learningtorankLow algorithms to learn the optimal weights of the different components using a separate dataset for training, we have
\begin{align}
\log \condprob{\doc}{\query} = Z
                    &+ \sum_i \alpha_i \log \undercondprob{i}{\finalQuery}{\doccontent} \label{eq:prvtm-retrieval-score} \\
                    &+ \beta \log \condprob{\doctime}{\finalQuery,\corpus} \label{eq:prvtm-temporal-feedback} \\
                    &+ \gamma \log \sum_{\vertical \in \selected{\query}} \condprob{\doctime}{\query,\vertical} \cdot \condprob{\query}{\vertical} \label{eq:prvtm-vertical-temporal-feedback} \\
                    &+ \sum_j \delta_j \log \prob{\docmeta^j}, \label{eq:prvtm-metadata}
\end{align}
where $\condprob{\finalQuery}{\doccontent}$ is the retrieval score of the document, $\doc$, given the expanded query, $\finalQuery$.
Instead of using a single estimate of the lexical component, $\condprob{\finalQuery}{\doccontent}$,
more accurate results are obtained by combining multiple estimates provided by different retrieval models (\Cref{eq:prvtm-retrieval-score}).
The weights $\alpha_i$ indicate the confidence in each retrieval model's estimate.
Since query expansion is used, the documents used for temporal feedback (see \Cref{eq:prvtm-temporal-feedback}) are retrieved using the expanded query.
This additional temporal feedback feature according to the corpus, $\corpus$, itself, is added on top of the estimation of temporal relevance from the external collections introduced by \Cref{eq:prvtm-vertical-temporal-feedback}.
To account for different ways to estimate the importance of a document from metadata we introduce \Cref{eq:prvtm-metadata}.
Next, we discuss the relationship between each feature of the above ranking model and the formal model.
\begin{table}[thpb]
	\caption{\learningtorankCap features.}
	\label{tab:prvtm-non-temporal-features}
	\centering
	\small
	\begin{threeparttable}
	\begin{tabular*}{0.85\linewidth}{@{\extracolsep{\fill}} l l @{}}
		\toprule
		Feature name & Feature description \\
		\midrule
		Doclen & Document length. \\
		\#URL & URL count. \\
		\#hashtags & Hashtags count. \\
		\#\@mentions & Mentions count. \\
		hasURL & 1 if it contains URL, otherwise 0. \\
		hasHashtags & 1 if it contains Hashtags, otherwise 0. \\
		hasMentions & 1 if it contains Mentions, otherwise 0. \\
		isReply & 1 if it is a Reply, otherwise 0. \\
		\#statuses & Total number of posts. \\
		\#followers & Total number of followers. \\
		\bottomrule
	\end{tabular*}
	\footnotesize
	\begin{tablenotes}
		\item \emph{Used in the \learningtorankLow methods, including \TERMEKDE.}
	\end{tablenotes}
	\end{threeparttable}
\end{table}

\paragraph{\learningtorankCAP Features}
The proposed model is composed of four main components that capture different aspects of search relevance in time-sensitive collections.
First, we employ three different retrieval models to obtain textual matching scores, \Cref{eq:prvtm-retrieval-score}.
They are, the \querylikelihoodLow retrieval model with Dirichlet prior smoothing (LM.Dir), BM25, and IDF.
Second, \Cref{eq:prvtm-temporal-feedback} includes a temporal feedback feature~\citep{efron_temporal_2014}, calculated over the documents retrieved from the main corpus $\corpus$ with the expanded query $\finalQuery$.

Third, the proposed model generalizes the integration of temporal evidence from external collections, \Cref{eq:prvtm-vertical-temporal-feedback}, aggregated into a single score.
In \Cref{eq:prvtm-vertical-temporal-feedback}, the importance of the publishing timestamp of the document $\doctime$ according to the external collections is estimated by a summation over the likelihood of each selected verticals.
For each vertical, $\condprob{\doctime}{\query,\vertical}$ returns the likelihood that an instant represented by $\doctime$ is relevant to the query $\query$ according to, $\RetrievedSet_\vertical(\origQuery)$.
The coefficients $\alpha_i$, $\beta$, $\gamma$, and $\delta_j$ correspond to the feature weights.
In contrast to previous work that often relies on a single source of temporal evidence,
\eg corpus, the proposed approach contemplates the use of several external collections.
The calculation of the temporal evidence feature over the documents retrieved from query-specific verticals can provide a more robust estimation of the relevant time periods for each query.

Fourth, many non-temporal and query-independent features \Cref{eq:prvtm-metadata}, were added
to improve effectiveness further,
such as quality features~\citep{choi_quality_2012}
and other commonly used features in \learningtorankLow
approaches to microblog search~\citep{xu_hltcoe_2014}. \Cref{tab:prvtm-non-temporal-features} lists the set of features.
This set of features captures microblog-specific information
that is useful for ranking such as,
number of statuses, number of followers,
number of URLs, and number of hashtags.
The number of words in the tweet was added as a feature to boost longer documents.

\subsection{Example: Temporal Evidence from External Collections}
\label{subsec:prvtm-example}

\begin{figure}[t]
	\centering
	\begin{minipage}{1\columnwidth}
		\centering
		\subcaptionbox{$\condprob{\instant}{\vertical}$.\label{fig:time-shard}}{
			\includegraphics[trim=20 5 0 5, clip=true, width=0.485\textwidth]{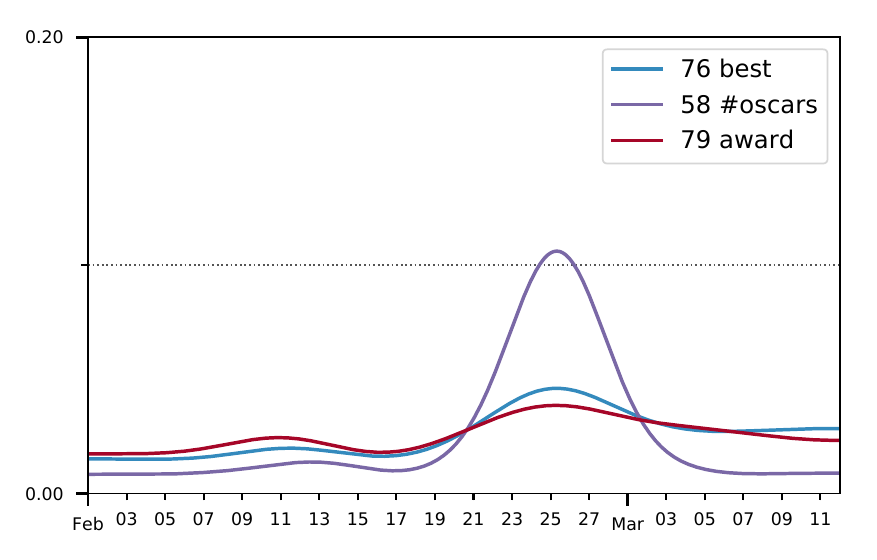}}
		\subcaptionbox{$\condprob{\instant}{\query,\vertical}$.\label{fig:time-fbshard}}{
			\includegraphics[trim=20 5 0 5, clip=true, width=0.485\textwidth]{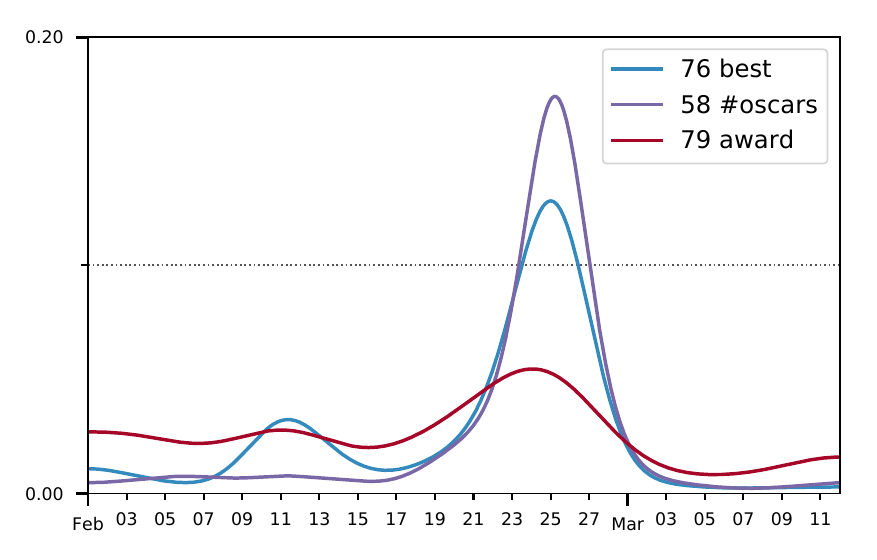}}
	\end{minipage}
	\caption{Temporal profiles of queries and collections.}
	\label{fig:temporal-evidence-shards-feedback}
	\vspace{-5mm}
\end{figure}

Let's examine an example in light of the described model.
Consider the 85th Academy Awards ceremony
that took place on February~24~2013,
at the Dolby Theatre in Hollywood, Los Angeles.
The top award winner, Argo,
winning the Oscar Award for \emph{Best Picture},
was a movie starring Ben Affleck.
This event sparked multiple processes on the Web,
such as the dissemination of news articles about the event,
and discussions and commentary on Twitter, \Cref{fig:temporal-evidence-shards-feedback}.
A person interested in surveying the general commentary and opinions
could procure a list of relevant accounts
to monitor posts in real-time.
However, journalists and other searchers
would most likely use a search engine 
to find more general information outside one's circle
about specific aspects of the event.

We dissected how multiple accounts
from news outlets and other verified account on Twitter
organized into different topical shards
can be used in the search process.
Using the TREC Microblog query MB195 - \enquote{Argo wins Oscar},
we plot two graphs
that show the temporal distribution of results
at the different stages of the framework.
Firstly, in \Cref{fig:time-shard}
we show the estimation of relevant time periods
using only the global statistics of each shard, via kernel density estimation
over the timestamps of all of its documents.

In this example we use a resource selection algorithm~\citep{aly_taily_2013}
to select the three most useful shards for the query.
All three topical shards selected exhibited
a larger probability around the time of the live broadcast.
Identified by the word \enquote{\#oscars}, Shard 58,
is relatively more bursty than the others.
Secondly, since topical shards are too broad we can fine-tune the estimation of the relevant time periods for a given query by finding a further subset of documents that are related to the query.
In \Cref{fig:time-fbshard}, we improved the estimation by searching over the same topical shards selected by the resource selection algorithm and for each one using for estimation the documents retrieved by the query \enquote{Argo wins Oscar}.
Shard 79 identified by the word \enquote{award} seems to be less focused than the others. 
The key insight from this comparison is that two topical shards have the most useful temporal information.

\section{Experimental Methodology}
\label{sec:study}
This section presents the evaluation of the methods described in the previous sections on the TREC microblog search test-bed.
In the TREC Microblog track problem of retrospective ad hoc retrieval, the user wishes to find the most up-to-date and relevant posts.
The task can be summarized as: at time $\instant$, find \emph{tweets} about topic $\query$.
Therefore, systems should favor highly informative \emph{tweets} relevant to the query topic that were published before the query time.

\subsection{Protocol}

Our experiments delve into the problem of re-ranking \emph{tweets} sampled using a standard retrieval method (\ie \querylikelihoodLow model) taking into account temporal crowd signals from different sources. 
In our experiments, we follow TREC and report the MAP and P30 results.
Statistical significance of effectiveness differences are determined using two-sided paired \textit{t}-tests following \citet{sakai_statistical_2014}.

\textbf{Filtering Duplicates and Languages.}
In the collection used, \emph{retweets} are considered not relevant because they are seen as duplicate documents.
Therefore, we filtered \emph{Twitter-style} \emph{retweets} using the tweet metadata available, and we also filter out \emph{RT-style} \emph{retweets} that start with \emph{RT}.
Moreover, assessors evaluated only relevant \emph{tweets} written in English, therefore we use the language filter \mbox{ldig}\footnote{\url{https://github.com/shuyo/ldig}} to remove \emph{tweets} in other languages.

\subsection{Datasets}
\label{subsec:tweets2013}

\subsubsection{TREC Microblog}
The Tweets2013 dataset is the most comprehensive evaluation resource for \adhoc retrieval on social media to date.
The Tweets2013 corpus is much larger ($\approx$240 million \emph{tweets}) than Tweets2011 (16 million \emph{tweets}) used in TREC 2011 and TREC 2012.
It was created by crawling Twitter's public sample stream over the period spanning from 1~February 2013 -- 31~March 2013.
The experiments were performed using both the query topics for the 2013 and 2014 editions of the TREC Microblog track~\citep{lin_overview_2013, lin_overview_2014}.
NIST provided relevance judgments TREC 2013 (60) and TREC 2014 (55) on a three-point scale: not relevant, relevant, and highly relevant.

\subsubsection{External Collections: Twitter Verified Accounts}
\label{subsec:verified}

We crawled the timelines of Twitter's verified users
($\sim$205k accounts as of Aug 2016)
collecting tweets
from the period 1 February -- 31 March 2013,
which matches the period covered
by the Tweets2013 TREC microblog dataset.
Twitter's verified accounts
belong to news organizations,
mass media,
and celebrities,
so the posts have higher quality
than a randomly sampled accounts.
The cleaner vocabulary also allows
the identification of interesting clusters more easily.

Topping the list of verified users (sorted by number of followers),
there are a number of singers, actors, and other celebrities.
Additionally, some accounts belong to companies that provide customer support through Twitter.
These accounts provide customer support using private messages sent via Twitter Direct Messages (DMs).
To be able to send DMs on Twitter, users have to follow each other.
Thus, to help remove these two types of unwanted accounts,
we extract two additional metrics for each account:
\begin{itemize}
\item the average number of tweets per day and
\item the ratio between the number of replies and total posts.
\end{itemize}

To select high quality informative sources we remove accounts that meet the following criteria:
$posts/day < 10$
and
$\frac{replies}{posts} > \rfrac{1}{3}$.
Accounts that belong to news media outlets and other mass media organizations,
typically produce a high volume of posts daily.
Thus, we remove accounts that have a low daily average number of posts (\eg @katyperry, @justinbieber, \etc).
News accounts and broadcasters seldom reply to other users on Twitter,
while accounts used by companies to provide customer support have a high ratio of replies (\eg @XboxSupport, @AppleCare, \etc).

Each account's timeline is then classified in terms of written language by sampling their five most recent posts
using ldig to remove non-English accounts.
A total of 645 accounts were used, totaling approximately 800k \emph{tweets}.

Tweets are tokenized using
Twokenize\footnote{\url{https://github.com/myleott/ark-twokenize-py}},
initially published alongside TweetMotif~\citep{oconnor_tweetmotif_2010}.
Preprocessing included removing URLs, email addresses,
numbers, times, \@mentions, and emoticons.
The tweets corpus was partitioned
using mini-batch \kmeansCap,
with the number of clusters empirically set to $K = 200$
since the corpus covers a large period of 2 months.

\subsection{Baselines and Experimental Systems}
\label{subsec:baselines}

\paragraph{Relevance baselines.}
The first baseline is the \querylikelihoodLow retrieval model with Dirichlet prior smoothing~\citep{zhai_study_2004} with $\mu = 2500$, which we will refer to as the \textbf{LM.Dir} model.
The second strong baseline, \textbf{LTR}, is a \learningtorankLow model combining multiple retrieval models (\ie LM.Dir, BM25, IDF) and the features in \Cref{tab:prvtm-non-temporal-features}.

\paragraph{Temporal baselines.}
There are three temporal ranking baselines: \textbf{Recency}~\cite{li_timebased_2003}, and \textbf{KDE(score)} and \textbf{KDE(rank)}~\cite{efron_temporal_2014}, two different variants of a state-of-the-art temporal feedback method.

\paragraph{Experimental systems.}
The \textbf{\EKDE} method consists in performing temporal feedback on external collections as described in \Cref{sec:temporal_vertical_feedback}.
The \textbf{\ERM} method uses the external collections, described in the previous section, to expand the initial query before searching the main corpus.
The \textbf{\TERM} experiment system uses time-based term expansion introduced in \Cref{sec:etbrm}.
Finally, the proposed \textbf{\TERMEKDE} experimental system uses both temporal vertical feedback and time-based term expansion.
Whenever KDE is used, we opted for the KDE(rank) variant due to its better performance on previous publications.
The \learningtorankLow methods use \textit{coordinate ascent} to optimize mean average precision (MAP).

\section{Results and Discussion}
\label{sec:prvtm-results}
In this section we start by comparing the retrieval results of the different baselines, temporal methods, and the experimental systems, and then present a qualitative analysis of the temporal distribution of the results retrieved by different systems.

\subsection{Retrieval Results}

\subsubsection*{\textbf{Time-based Relevance Models Using External Collections}}
\label{subsec:temporalrelevancemodels}

\begin{table*}[t]
\caption{TREC evaluation results.}
\subcaptionbox{TREC 2013 dataset results.\label{tab:prvtm_temprank_all_2013}}{
	\begin{threepartsubtable}
		\begin{tabularx}{0.425\textwidth}{l *{3}{Y}}
			\toprule
			Method & \multicolumn{1}{l}{MAP} & \multicolumn{1}{l}{P30}        & \multicolumn{1}{l}{Rprec} \\ \midrule
			LM.Dir                & 0.2629 & 0.4622 & 0.3094 \\
			Recency               & 0.2663 & 0.4611 & 0.3115 \\
			KDE(score)            & 0.2583 & 0.4517 & 0.3004 \\
			KDE(rank)             & 0.2736\wc & 0.4878\wc & 0.3178\wc \\
			LTR					  & 0.2787 & 0.4617 & 0.3193 \\
			\midrule
 			\ERM                  & 0.2797 & 0.4528 & 0.3167 \\
			\TERM                 & 0.2824\wc & 0.4700 & 0.3233 \\
			\EKDE			      & 0.2889\bc & \textbf{0.5061}\bc & \textbf{0.3322}\bc \\
			\TERMEKDE             & \textbf{0.2900}\wc & 0.4850 & 0.3229 \\
			\bottomrule
		\end{tabularx}
		\footnotesize
 		\begin{tablenotes}
			\item \emph{Symbols \textnormal{\wc} and \textnormal{\wl} stand for a $p < 0.05$ statistical significant improvement over KDE(score) and LTR respectively (\textnormal{\bc} and \textnormal{\bl} for $p < 0.01$).}
 		\end{tablenotes}
	\end{threepartsubtable}
}
\qquad
\subcaptionbox{TREC 2014 dataset results.\label{tab:prvtm_temprank_all_2014}}{
	\begin{threepartsubtable}
		\begin{tabularx}{0.425\textwidth}{l *{3}{Y}}
			\toprule
			Method & \multicolumn{1}{l}{MAP} & \multicolumn{1}{l}{P30}        & \multicolumn{1}{l}{Rprec} \\ \midrule
			LM.Dir                & 0.4316 & 0.6315 & 0.4552 \\
			Recency               & 0.4323 & 0.6382 & 0.4576 \\
			KDE(score)            & 0.4205 & 0.6303 & 0.4476 \\
			KDE(rank)             & 0.4399 & 0.6406 & 0.4664 \\
			LTR					  & 0.4469 & 0.6721 & 0.4625 \\
		    \midrule
			\ERM                  & 0.4705 & 0.6394 & 0.4890 \\
			\TERM                 & 0.4738\bc & 0.6442 & 0.4927\wc \\
            \EKDE			      & 0.4643\bc\bl & 0.6776\bc & 0.4869\wc\bl \\
			\TERMEKDE             & \textbf{0.5183}\bc\bl & \textbf{0.6970}\wc & \textbf{0.5138}\bc\bl \\
			\bottomrule
		\end{tabularx}
		\footnotesize
 		\begin{tablenotes}
			\item \emph{Symbols \textnormal{\wc} and \textnormal{\wl} stand for a $p < 0.05$ statistical significant improvement over KDE(score) and LTR respectively (\textnormal{\bc} and \textnormal{\bl} for $p < 0.01$).}
 		\end{tablenotes}
	\end{threepartsubtable}
}
\end{table*}

In this section we analyze the influence of time-based relevance models. 
The organization of the expansion corpus into topic-based verticals
makes the query expansion process \textit{temporally focused}.
Verticals created by a partitioning algorithm using a topic-based similarity criteria exhibited different temporal profiles.
The distribution of documents contained in each topic-based vertical is biased towards the time periods for when the vertical is most relevant.
Following the temporal cluster hypothesis, the temporal relevance estimate extracted
using the timestamps from the verticals selected was integrated into the retrieval process.
In the \prfLow term selection stage it is used to generate \textit{temporally focused} query expansion terms.
In \Cref{tab:prvtm_temprank_all_2013} and \Cref{tab:prvtm_temprank_all_2014} we present a comparison of the results of MAP and P30 in the TREC 2013 and 2014 test topics.
By estimating the relevance models using the proposed time-sensitive term selection approach (\TERM), the retrieval effectiveness always improved against the non-temporal method (\ERM).
In fact, in TREC 2013 we observe a large effect of time-sensitive term selection on P30 when using the proposed vertical feedback architecture.
Overall, we found that time-sensitive term selection is effective when used in standard pseudo-relevance feedback as well as in the proposed vertical feedback architecture.

\subsubsection*{\textbf{Estimating Temporal Relevance Using External Collections}}
\label{subsec:prvtm-tempestimation}
In this section we analyze the importance of temporal feedback from external collections. 
The major difference between \EKDE and \TERMEKDE is that the former uses the vertical feedback architecture for temporal feedback only, while the latter uses this architecture for query expansion via a time-aware pseudo-relevant vertical feedback method.
In addition, it uses the estimate of temporal relevance obtained from temporal feedback on documents retrieved from the corpus using the expanded query.
Like the LTR method, \EKDE is based only on the re-ranking of the documents retrieved by an initial retrieval method (\ie LM.Dir).
It is, therefore, very interesting that the \EKDE is not only very competitive against LTR and the KDE-based methods, but also with \TERMEKDE.
In the TREC 2013 queries, \EKDE even outperformed \TERMEKDE for both top-precision metrics, P30 and Rprec.
\TERMEKDE outperformed the other methods on MAP, but the difference was not statistically significant against \EKDE.
In the TREC 2014 queries the \TERM-based methods outperform \EKDE on the recall-oriented metrics, MAP and Rprec.
\TERMEKDE statistically significantly outperformed \EKDE in the recall-oriented metrics, MAP and Rprec, in part due to the use of the \TERM method in \TERMEKDE to obtain the candidate set of documents for re-ranking. 

\subsubsection*{\textbf{Full Model Analysis}}
\label{subsec:prvtm-effectiveness-all}
To conclude the retrieval results analysis, we examine the overall gains offered by temporal evidence from topic-based external collections.
The results of the evaluation on the two TREC test datasets
are summarized in~\Cref{tab:prvtm_temprank_all_2013} and~\Cref{tab:prvtm_temprank_all_2014}.
We present the results for three retrieval effectiveness metrics: MAP, P30, and Rprec.
We found that \TERMEKDE can outperform non-temporal
\learningtorankLow models as well as state-of-the-art temporal ranking methods.

The \TERMEKDE method statistically significantly outperforms KDE(score) in both sets of queries.
MAP improved 12.3\% and 23.3\% in the TREC 2013 and TREC 2014 topics respectively.
Additionally, for the TREC 2014 topics the MAP result improved 17.8\% over KDE(rank) and was statistically significant.
Although in terms of P30, \TERMEKDE did not outperform KDE(rank) for the TREC 2013 queries, it outperformed KDE(score) albeit the result was not a statistically significant.
In contrast, the improvements on P30 with \TERMEKDE on the TREC 2014 topics reached a statistically significant result of 10.6\% over KDE(score) and 8.8\% over KDE(rank), respectively.

\TERMEKDE outperforms the LTR baseline consistently across all metrics on both sets of queries.
The improvements of \TERMEKDE in MAP and Rprec over LTR in the TREC 2014 topics were statistically significant, 16.0\% and 10.0\% for MAP and Rprec, respectively.

\begin{figure*}[h]
	\centering
	\subcaptionbox{MB124 -- \enquote{celebrity DUI}\label{fig:prvtm-MB124}}{
		\includegraphics[trim=5 0 10 0, clip=true, width=0.244\textwidth]{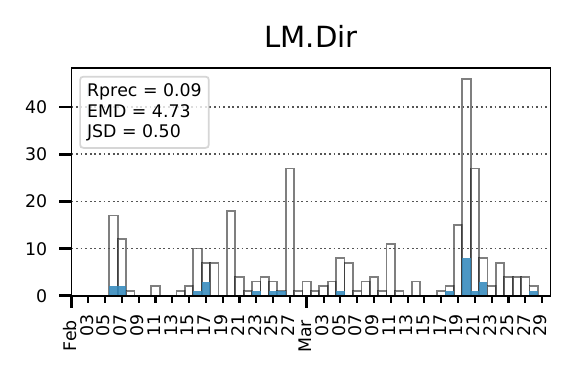}
		\includegraphics[trim=20 0 10 0, clip=true, width=0.22\textwidth]{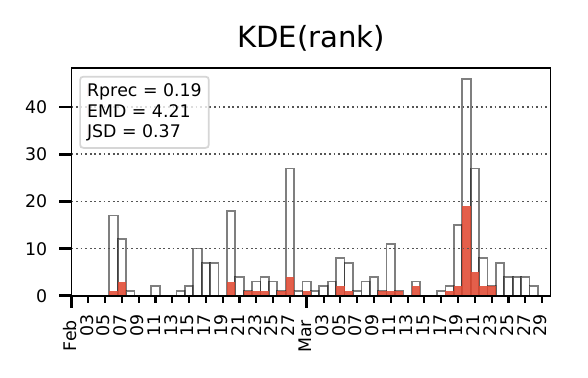}
		\includegraphics[trim=20 0 10 0, clip=true, width=0.22\textwidth]{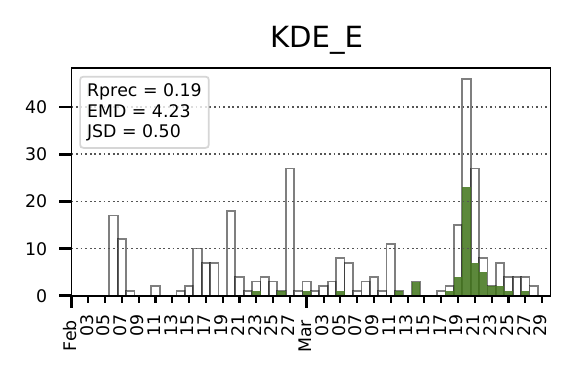}
		\includegraphics[trim=20 0 10 0, clip=true, width=0.22\textwidth]{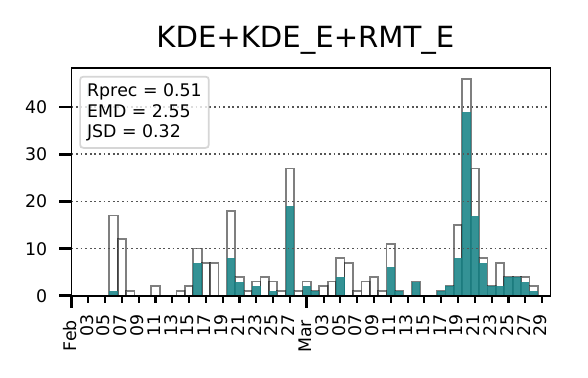}
	}
	\\
	\subcaptionbox{MB133 -- \enquote{cruise ship safety}\label{fig:prvtm-MB133}}{
		\includegraphics[trim=5 0 10 0, clip=true, width=0.244\textwidth]{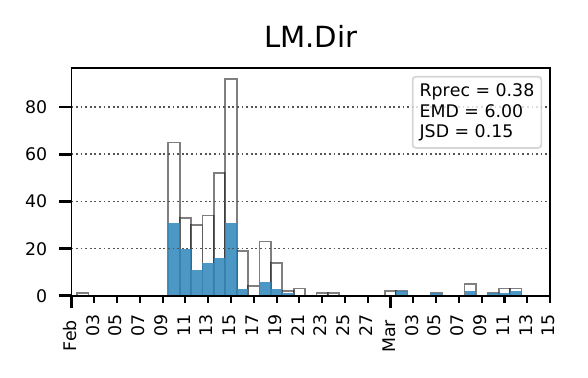}
		\includegraphics[trim=20 0 10 0, clip=true, width=0.22\textwidth]{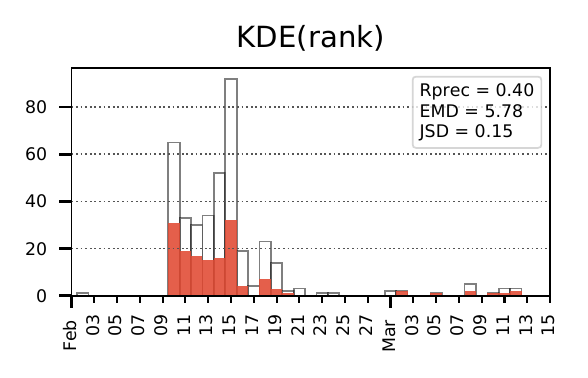}
		\includegraphics[trim=20 0 10 0, clip=true, width=0.22\textwidth]{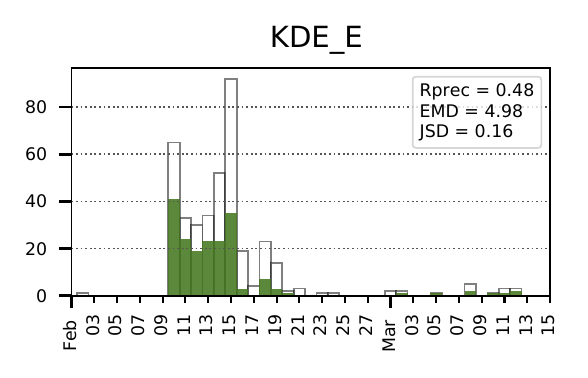}
		\includegraphics[trim=20 0 10 0, clip=true, width=0.22\textwidth]{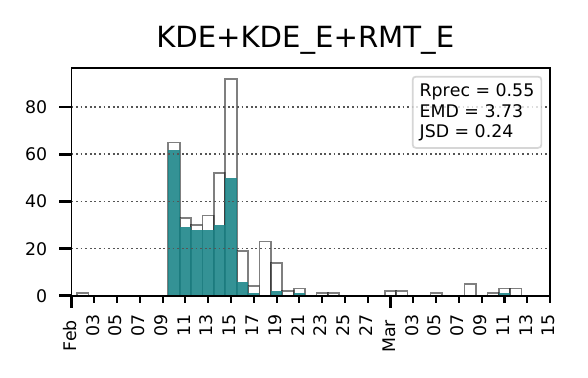}
	}
	\\
	\subcaptionbox{MB178 -- \enquote{Tiger Woods regains title}\label{fig:prvtm-MB178}}{
		\includegraphics[trim=5 0 10 0, clip=true, width=0.244\textwidth]{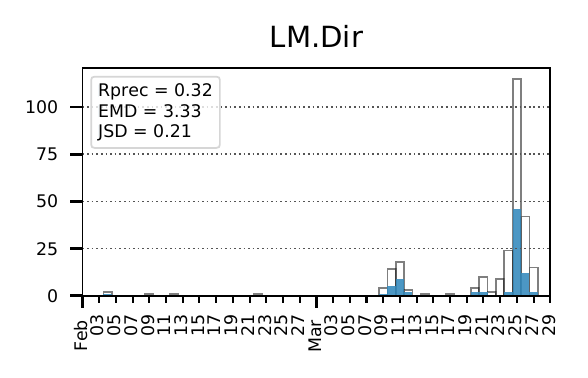}
		\includegraphics[trim=20 0 10 0, clip=true, width=0.22\textwidth]{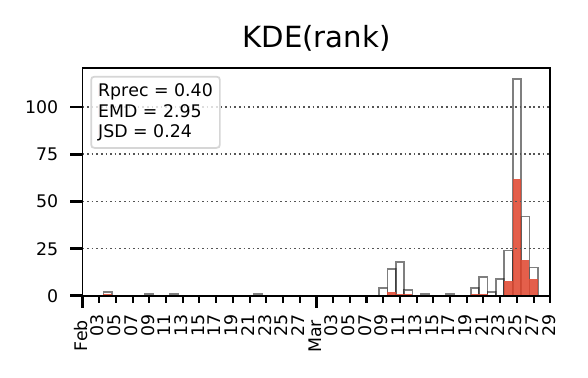}
		\includegraphics[trim=20 0 10 0, clip=true, width=0.22\textwidth]{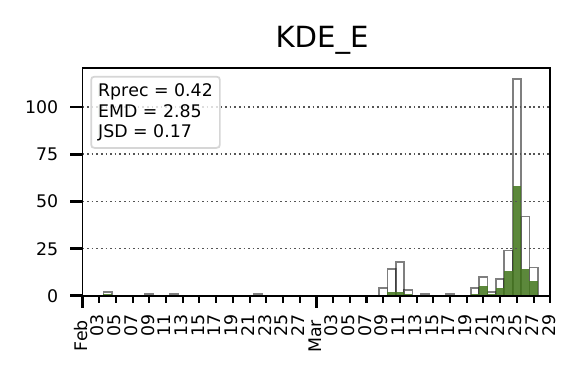}
		\includegraphics[trim=20 0 10 0, clip=true, width=0.22\textwidth]{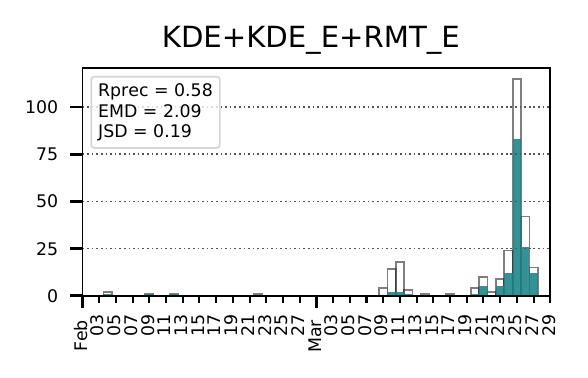}
	}
	\\
	\caption[Temporal profiles of queries and fit to the true distribution.]{Temporal profiles of queries and fit to the true distribution. The colored area of the bars represents the portion of relevant documents retrieved at a depth of $\relvar$, where $\relvar$ is the number of relevant documents in the ground truth (\ie Rprec).
	}
	\label{fig:deconstructed-topics}
\end{figure*}

\subsection{Temporal Distribution Analysis}
\label{subsec:jsdanalysis}

This section aims to provide extra insights to understand the different performance of the retrieval methods in light of the effect on the temporal distribution of their top ranked documents.
With this objective in mind, we look into a temporal representation of the \textit{R-Precision} metric,~\Cref{fig:deconstructed-topics}:
we plot the ground-truth distribution of the $R$ relevant documents of each query (empty bars) against the relevant documents retrieved at rank depth $R$ (shaded bars). A perfect method retrieves only relevant documents, hence completely filling the empty bars. This visualization allows us to see if the methods are returning documents from the time periods that contain more relevant documents in the ground-truth.
The plotted methods include the LM.Dir (no temporal evidence), KDE(rank) (temporal evidence from the corpus), \EKDE (temporal evidence from external collections), and \TERMEKDE (external, temporal feedback and time-based relevance model.
Additionally, we present the EMD metric to quantify the difference between the temporal distribution of the retrieved documents and the true distribution. It is interesting to observe the direct relation between the EMD and Rprec results.

In \Cref{fig:deconstructed-topics}, we plot some topics that improved the most.
For all the queries shown, we observe that the temporal distribution of the top documents agrees with the temporal distribution of the documents in the ground truth.

For the top performing topic (see~\Cref{fig:prvtm-MB124}) we can see that \EKDE 
retrieves
documents from the most relevant time period.
However, with \TERMEKDE by using temporal query expansion, additional relevant time periods are found and retrieved.
We can see that \TERMEKDE seems to retrieve more documents from the most relevant time period but it retrieves some documents from this second time period as well.

In the case of topic 133 \enquote{cruise ship safety}, \Cref{fig:prvtm-MB133}, it is clearly visible that \TERMEKDE is able to focus its retrieval towards documents published in February 10 and the following week.
Inspecting the documents we found mentions to the Carnival Triumph cruise ship incident.
This cruise ship set sail on February 7 and three days later (February 10) suffered an engine room fire.

The temporal distribution of the ground truth for topic 178 \enquote{Tiger Woods regains title},~\Cref{fig:prvtm-MB178}, indicates that most of the relevant documents are near the time of the query.

LM.Dir follows the temporal distribution of the ground truth.
Nevertheless,
the temporal distribution of the documents retrieved using \EKDE and \TERMEKDE shows that they can retrieve more documents from the most relevant days.

\section{Conclusions}
\label{sec:conclusion}
This paper presented the \TERMEKDE
a time-aware and topic-aware pseudo-relevance feedback framework
that mines textual and temporal signals
from multiple information sources on Twitter.
It explores the signals from verified accounts posts on Twitter,
and temporal feedback to estimate the temporal relevance of search topics.
The information streams from the verified accounts are
automatically partitioned into verticals according to their topic. 
\vspace{-5mm}
\paragraph{Time-aware topical-based evidence mining.}
The results of the experiments confirmed our hypothesis
that jointly modeling the topicality and temporality
improves the estimation of relevance models,
and yields improvements in Rprec along the timeline.

\paragraph{Efficient use of external collections.}
Building on recent advances, we show how to exploit the temporal heterogeneity of multiple external information verticals for time-aware ranking. These topic-based external verticals are exploited at two stages of the retrieval process: query expansion and time-aware ranking.

\begin{acks}
This work has been partially funded by the \grantsponsor{CMUP}{CMU Portugal}{} research project GoLocal Ref. \grantnum{CMUP}{CMUP-ERI/TIC/0033/2014}, by the \grantsponsor{EU H2020}{H2020 ICT}{} project COGNITUS with the grant agreement n\textsuperscript{o} \grantnum{EU H2020}{687605} and by the \grantsponsor{FCT}{FCT}{} project NOVA LINCS Ref. \grantnum{FCT}{UID/CEC/04516/2013}.
\end{acks}


\end{document}